\documentclass[aps,physrev,preprint,groupedaddress]{revtex4-2}

\usepackage{lipsum}
\usepackage{graphicx}
\usepackage{textcomp}
\usepackage{bm}
\usepackage{marvosym}
\usepackage{amssymb}
\usepackage{lipsum}
\usepackage{labelfig}
\usepackage{comment}
\usepackage[dvipsnames]{xcolor}

\begin{document}


\title{Stress network dynamics influence on large particle segregation}


\author{Alexander Navarrete$^{1}$, Leonardo Gordillo$^{2}$ and Tomás Trewhela$^{1}$}
\email{alenavarrete@alumnos.uai.cl}
\affiliation{$^{1}$Facultad de Ingeniería y Ciencias, Universidad Adolfo Ibañez, Av. Padre Hurtado 750, 2562340 Viña del Mar, Valparaíso, Chile\\
$^{2}$Departamento de Física, Universidad de Santiago de Chile, Estación Central, Santiago 9170124, Chile}


\date{\today}

\begin{abstract}
A plethora of natural and industrial shear-driven granular flows exhibit particle-size segregation. Its occurrence is commonly attributed to two primary mechanisms: kinetic sieving and squeeze expulsion. While kinetic sieving is relatively well understood, squeeze expulsion lacks a clear mechanical explanation and direct experimental evidence due to difficulties in measuring stresses in granular media. Here, we investigate force networks around a large intruder in a bidimensional granular shear cell. We use transparent, birefringent disks to visualize stress chains via photoelasticity. Experiments were conducted with two different granular media to study force chains over size ratios between the intruder and surrounding particles of 1.25 to 4.0. Particle Tracking Velocimetry and G² analysis are used to quantify particle trajectories and identify active grains. These methods enable us to measure force-chain lengths and structures around the intruder through the gap factor. Our results confirm that squeeze-expulsion strongly depends on stress transmission. Larger size ratios lead to longer force chains and greater particle participation in the global stress network. In parallel, stress fluctuations predominate in driving or restraining intruder motion by forming anisotropic force chains. These findings advance the understanding of granular segregation by clarifying the link between force-network dynamics and segregation mechanics.
\end{abstract}


\maketitle

\section{Introduction}
\label{sec:intro}

Shaking or shearing can fully reconfigure a disordered mixture of particles of different sizes into an ordered distribution, with smaller and larger particles accumulating at the bottom and the surface, respectively. Many granular systems in industrial, natural, and space environments likewise exhibit particle-size segregation when subjected to shear \cite{Gray18,Umbanhowar19,Thornton26}. Such gravity- and shear-driven particle rearrangement is the result of two main mechanisms described conceptually as kinetic sieving \cite{Middleton70} and squeeze expulsion \cite{Savage88}.
Kinetic sieving occurs when particle layers act as sieves through which small particles flow through. Squeeze expulsion happens when smaller particles push larger ones as the granular flow tighten the gaps. This latter mechanism remains poorly understood, with its origin being a subject of much debate in the past decade \cite{Thomas00,Jing17,vanderVaart18,Staron18,Kiani21,Trewhela21b,vanSchrojenstein21}.

The migration of large particles to the surface of granular materials under mechanical forcing \cite{Williams68} is widely recognized in the literature as the \textit{Brazil-nut effect} \cite{Rosato87,Ottino00}, and can be triggered by both vibration and shearing. But in the specific case of shear, particle-size segregation is sustained by both kinetic sieving and the squeeze expulsion mechanisms \cite{Savage88,Gray18}. Despite the various monikers and qualitative descriptions, a key question remains unanswered: why do large particles segregate? As a clue, there are, some well-known conditions that need to be met for large grains to rise. Many authors agree on the need for voids or local dilation, friction, and shear rate transmission \cite{Rosato87,Guillard16,Jing17,vanderVaart18,vanSchrojenstein21,Trewhela21b,Tripathi21}. Yet, the uncertainty arises when discussing the origin of large-particle segregation: drag, lift, or buoyancy forces, or a normal force imbalance, appear as plausible causes \cite{Guillard14,Jing20,Kiani21}. Guillard et al. \cite{Guillard16} proposed a scaling law for the segregation force acting on a large intruder and suggested a lift force that depends on the local stress distribution. Likewise, van der Vaart et al. \cite{vanderVaart18} developed the concept of a buoyancy-like force within granular flows, drawing an analogy with viscous drag in fluids. In another approach, Staron \cite{Staron18} argued that the primary driver of large particle segregation is the fluctuation of contact forces and the anisotropic structure of force chains, which intermittently break down and rebuild, thereby producing a net upward thrust when there is a force imbalance. These studies underscore that the upward migration of large particles through squeeze-expulsion is influenced by a complex interplay of stress transmission, particle size ratio, force network dynamics, and local structural heterogeneity.

To address such a complex interplay, it is key to measure granular-flows force distribution, which is challenging due to real grains opacity. To overcome this hurdle, it is possible to employ birefringent grains in experiments; inferring the acting stresses using photoelasticity following recent advancements on stress fluctuations and network dynamics \cite{Drescher72,Howell99,Daniels17,McMillan25}.

In this article, we shed light on the relation between the segregation of a large particle with a different diameter, i.e. an intruder, from that of the surrounding granular medium and stress network dynamics. In \S \ref{sec:methods}, we present the bidimensional oscillating shear cell used for the experiments and the photoelasticity technique. Next, a validation of the segregation scaling law of \cite{Trewhela21a,Trewhela21b} is conducted to determine the segregation parameters for the polyurethane disks used as grains. Finally, in \S \ref{sec:results}, we present the probability density functions $PDF$ for both the stress chain lengths and the networks' gap factor. We found that these functions are strongly dependent on the particle-size ratio, showing that large particle-size anisotropy influences the characteristics of the surrounding stress network, thus resulting in its own segregation. Our work aims to bridge stress transmission and granular segregation through a novel approach and elucidate the physical mechanisms behind large grain segregation. Besides the peculiar observation of large particles at the surface in everyday granular materials like nuts \cite{Rosato87} or chips, we emphasize the crucial role of large particle segregation in determining debris flow runouts, avalanche levees, heterogeneous mortars and asphalts, and even discarding pharmaceutical batches. Therefore, our understanding of the mechanics behind large particle segregation can help to improve predictive tools for the proper handling and modeling of granular materials.

\section{Experimental methods and techniques}
\label{sec:methods}
\subsection{Bidimensional shear cell setup}

\begin{figure}
\begin{center}
\SetLabels
\L (-0.005*0.98) ($a$)\\
\L (0.66*0.98) ($b$)\\
\L (-0.005*0.37) ($c$)\\
\L (0.38*0.95) Crank\\
\L (0.16*0.95) Plunger\\
\L (0.03*0.95) Panels\\
\L (0.405*0.76) Polarizing filters\\
\L (0.405*0.64) PVC side levers\\
\L (0.405*0.52) Sliding  base\\
\L (0.62*0.84) \color{white}{$\omega$}\\
\L (0.74*0.66) \color{white}{$h$}\\
\L (0.79*0.73) \color{white}{$d_{s}$}\\
\L (0.85*0.69) \color{white}{$d_{l}$}\\
\L (0.925*0.57) \color{white}{$\theta$}\\
\L (0.87*0.47) \color{white}{$W$}\\
\endSetLabels
\strut\AffixLabels{\includegraphics[width=\textwidth]{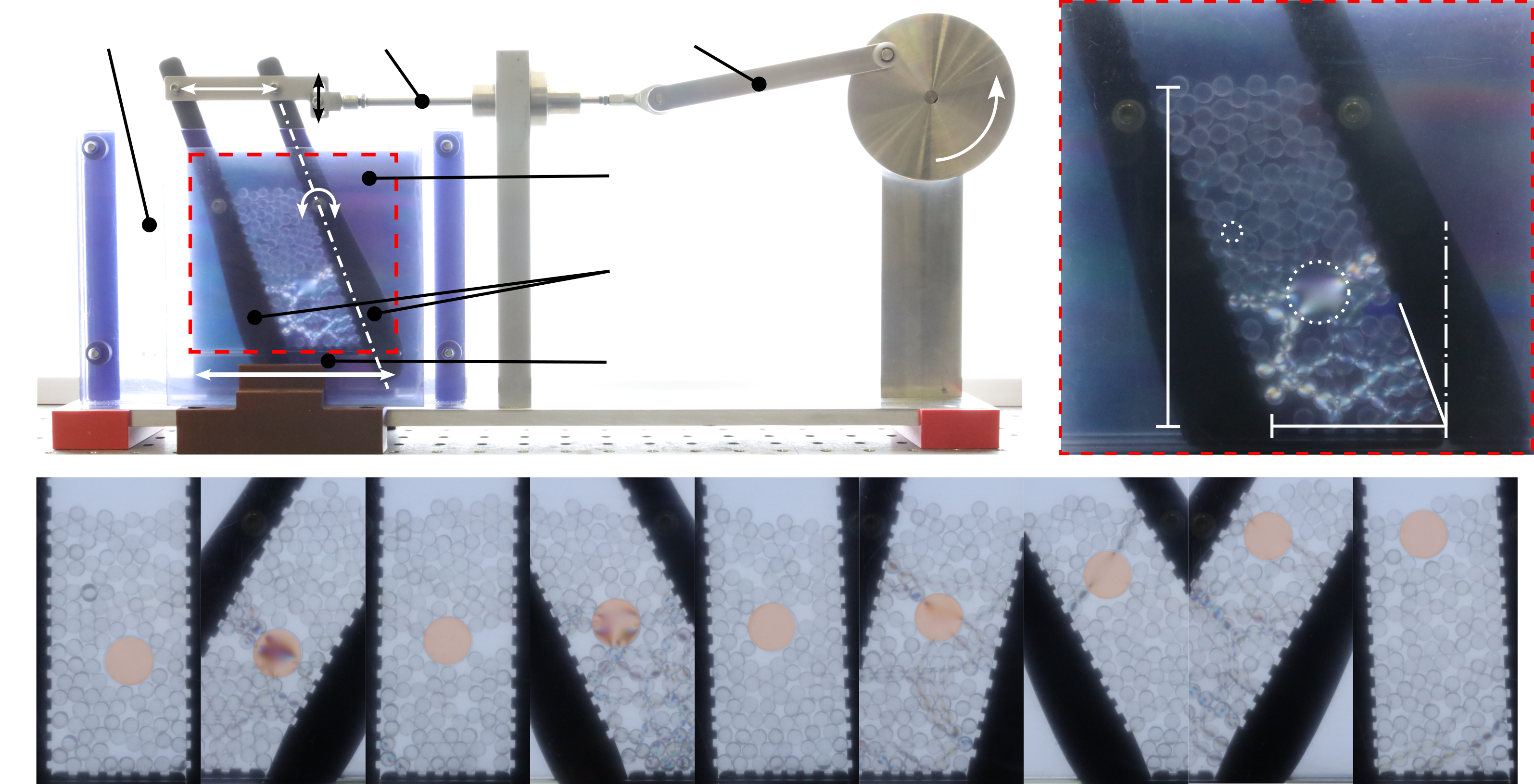}}
\end{center}
 \caption{($a$) The bidimensional oscillatory shear cell used in the experiments. A stepper motor rotates a wheel with frequency $\omega$, which actuates a crank-plunger mechanism that moves the parallel PVC levers horizontally. As the PVC levers shear the material within the cell, two polarizing filters placed at the front and back of the cell allow stress network visualization. ($b$) Close up of the sheared granular material. A clear stress network is observed with the help of photoelasticity.  ($c$) Image sequence of the large particle (intruder in pale orange to ease visualisation) segregating (rising) to the surface due to the action of the oscillating shear. A supplementary movie (Movie 1) shows the operation of the shear cell.}
 \label{fig:setup}
\end{figure}
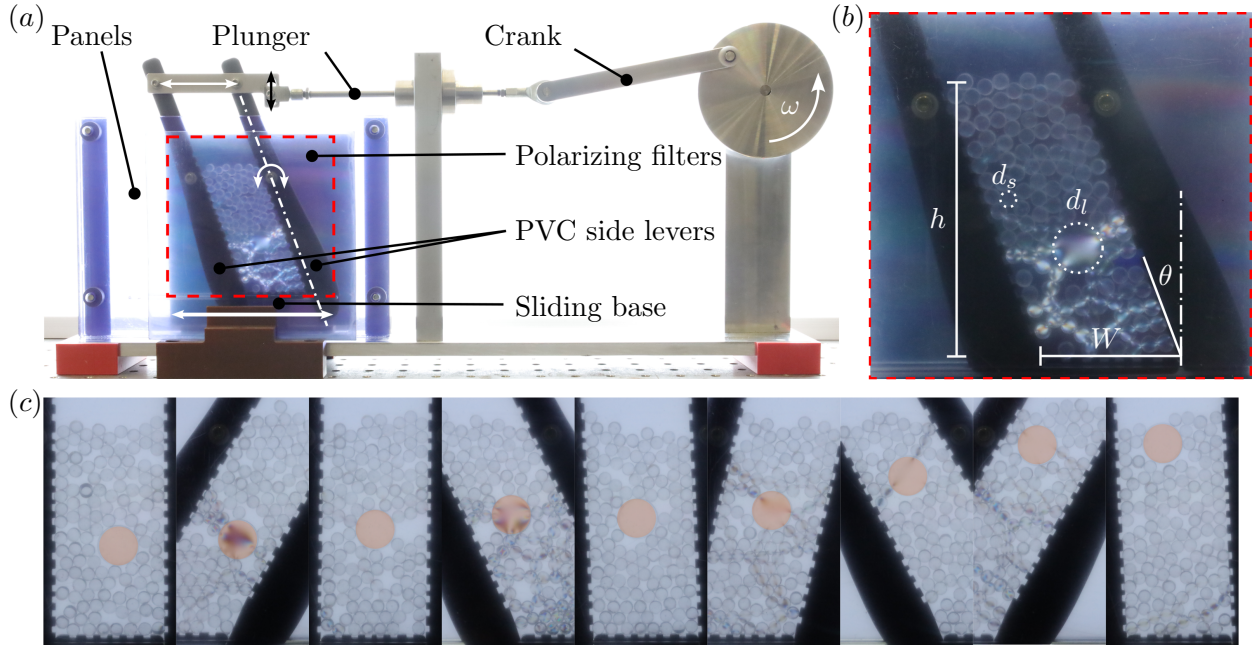

The experimental setup consists of a ($W\times h\times \epsilon$) $45 \times 90 \times 5\mathrm{mm^3}$-gap  Hele-Shaw cell built from polyvinyl chloride (PVC) with a sliding base. Two rigid PVC levers in its interior, separated by a distance $W = 45$ mm, pivot each on a pair of steel rod axes at their midpoint and top (figure \ref{fig:setup}\emph{a}). The axes and sliding bottom plate are respectively anchored and hinged at the cell front and back, keeping the levers parallel as they swivel and pull the sliding base. After the gap is filled with grains, the shear is ultimately produced by a stepper motor (Micro Motors E192.12.336) and a crank that simultaneously transmits the motion to a metal bar connecting the two levers.  A detailed image description of the setup is provided in figure \ref{fig:setup}\emph{a-b}. It is worth noting that the lever surface roughness is randomly distributed and scales with the diameter of the small particles, while the cell's front and back walls are transparent for visualization.

In this experimental configuration, the shear rate is simply $\dot{\gamma}_{e}(t)=\omega|\text{cos}(\omega t)|\text{tan}\,\theta_{\text{max}}$. During a cycle, this shear rate can be averaged to yield a cycle-averaged (mean) shear rate of $\bar{\dot{\gamma}}=\frac{2\omega}{\pi}\,\text{tan}(\theta_{\text{max}})$. Here, the angle $\theta$ corresponds to the inclination of the side plates during a cycle. It reaches a maximum value, $\theta_{\text{max}}=30$\textdegree, which is kept fixed in our experiments, as in previous studies using this configuration \cite{vanderVaart15,Trewhela21a,Trewhela21b}. During the experiments, the motor operates at a fixed voltage, resulting in a rotational period of $T = 3.6$ s. This gives a corresponding angular frequency of $\omega = 2\pi/T=1.745$ s$^{-1}$. As a result, a constant mean shear rate of $\bar{\dot{\gamma}}=0.271$ s$^{-1}$ was exerted in all our experiments. 

Width and shear rate were kept constant for this study, as the role of shear rate is well known for segregation in this configuration \cite{vanderVaart15,Trewhela21a}. Similarly, variation in width does not affect segregation dynamics as long as the granular medium is well sheared \cite{Trewhela21b}. Too narrow or too wide conditions are also known to produce singular effects, such as Jansen's \cite{Janssen95} or non-uniform shear transmission (in the form of shear banding), which are undesirable for studying segregation dynamics.

\subsection{Polariscope and experimental sets}

To visualize active stresses in the granular medium via photoelasticity, we used grains consisting of birefringent polyurethane disks, which produced color fringes in their interior when actively stressed (see figure~\ref{fig:setup}\emph{b-c}) \cite{Daniels17,Abed19,Ramesh20}. These disks were fabricated using the commercial polyurethane \textit{ClearFlex 50}, cast into silicone molds (\textit{Mold Star 15 Slow}) \cite{Photoelasticity24}. The disks were polymerized in various diameters:
$d_{s}=\lbrace 5,8\rbrace$ mm for small particles and
$d_{l}=\lbrace 10:2:20\rbrace$ mm for large intruders. Each $d_{s}$ medium, combined with all the $d_{l}$ intruders, yielded two experimental sets of 6 cases, for a total of 12 experiments. These two sets, with the combination of medium and intruder, and their resulting size ratio, are summarized in Table~\ref{tab:experimental-matrix}. Our matrix ensured a controlled size ratio $R$ that ranged from 1.25 to 4. To achieve a bulk of grains of constant height $h=90$ mm suitable for experiments, two typical samples of $d_s=5$-mm and $d_s=8$-mm particles required, respectively, 120 and 62 particles, plus an extra single intruder of diameter $d_l$.

We built a polariscope for photoelasticity stress visualization. The optical configuration consisted of background white illumination and two circular polarizing filters (right and left) placed in front of and behind the birefringent specimen in the shear cell. The polariscope configuration also included a Basler acA2000-165uc color camera equipped with a 12-mm C-Mount lens placed in front of the setup to acquire the stress network dynamics during the large particle segregation (figure \ref{fig:setup}\emph{c}). Our polariscope had a resolution of 1488 $\times$ 1056 pixels, with image acquisition rates of 62.5 fps. Both resolution and frequency were determined based on the chain and network duration, aiming to accurately capture their appearance and disappearance.

The experimental protocol was as follows: the oscillatory shear cell was first placed in the vertical position and filled with two rows of small particles, over which the intruder of diameter $d_{l}$ was placed. The cell was then filled with the remaining small particles to a height of $h=9$ cm. After being set, the rotating motor was powered up, shearing the grains that segregate the intruder. The camera recorded the dynamics until the intruder reached  the top of the cell. To finish the protocol, we emptied the cell to fully restore the original conditions to start the following experiment.

\begin{table}[ht]
\centering
\caption{Experimental size ratios $R=d_{l}/d_{s}$ considered for this study. Each experiment of a total of 12 is defined by the combination of the granular medium diameter $d_{s}$ and the intruder diameter $d_{l}$.}
\begin{tabular}{lcccccc}
\hline
\textbf{Medium diameter} & \multicolumn{6}{c}{\textbf{Intruder diameter $d_l$ (mm)}} \\ 
 $d_{s}$ mm & 10 & 12 & 14 & 16 & 18 & 20 \\ 
\hline
5  & 2.00 & 2.40 & 2.80 & 3.20 & 3.60 & 4.00 \\ 
8  & 1.25 & 1.50 & 1.75 & 2.00 & 2.25 & 2.50 \\ 
\hline
\end{tabular}
\bigskip
\label{tab:experimental-matrix}
\end{table}

\subsection{Particle tracking and contact identification}

The acquired images consisted of a collection of disks moving relative to each other. The first step in image processing was identifying the disk positions and trajectories. We employed a Hough transform algorithm \cite{Illingworth88} to detect the disks, along with their positions and diameters, based on their circular shape after image postprocessing. Once all particles were identified, a Brownian diffusion probability method combined with a Hungarian algorithm  tracked particles through the experiment and solved the assignment problem between detected particles to yield trajectories \cite{Kuhn95,Crocker96}. Particle identification and tracking serve as the foundation for subsequent stress chains or network identification, especially for those networks that involve the intruder.

From the experimental polarized images, it is possible to correlate the local light intensity with the principal stress difference $\Delta\sigma=\sigma_{xx}-\sigma_{zz}$, where $\sigma_{ii}$ are the principal stresses applied to the disks in the $x$ and $z$ directions. The correlation is based on the squared-gradient analysis method $G^{2}=\frac{1}{N}\Sigma_{i,j\in \text{ROI}}|\nabla I_{ij}|^{2}$, which takes the light-intensity gradient $\nabla I_{ij}$ for each $\lbrace i, j\rbrace$ pixel in the image, where $N$ is the total number of pixels within the ROI. Since $I=I_{0}\,\text{sin}^{2}\left(\epsilon\pi\,\Delta\sigma\,c/\lambda\right)$, the gradient provides the variation of the average intensity value within disks as the disks deform under shear, where $c=2430\cdot10^{-12}$ m$^{2}$/N is the photoelastic constant of \emph{ClearFlex 50}, $\epsilon=4$ mm is the disk thickness, and $\lambda_P$ is the wavelength of the light source used in the polariscope. Therefore, as the local stress distribution is encoded within the disk in the form of fringe patterns, stress can be determined from the very same changes in fringe light intensity variations.

Once the disk positions and internal stresses were available, contact detection was performed to determine the stress networks among disks. This processing step uses the PeGS modular code \cite{Lee25}, which first identifies contacts based on particle adjacency and then refines them by detecting high-$G^2$ zones at the edges of the particle. To finally classify particle connections as a chain, we constructed the contact network between all particle neighborhoods and applied the following criteria: if two particles were in physical contact and both classified as active (by having a $G^2$ above a prescribed threshold), they form a link in a chain. Yet, only networks with more than two contacts (three particles, $N_{p}\geq3$) were considered active networks, ensuring a minimal structural definition. A supplementary movie shows binarized images of the detected contacts and active stress networks (Movie 2).

\subsection{Stress network statistics}

Following stress and contact network identification, we choose a suitable statistical characterization of the identified chains for  postprocessing. In our experiments, spatial structure and topological organization were key for particle-size segregation. Single-parameter descriptors like the mentioned $G^2$ or chain length proved to be insufficient, having no connection to the size ratio or the particles' diameter. To overcome these limitations, we employed two parameters: the gap factor $g_c$ and the mean network order $L$.

The gap factor is defined as $g_{c}=1-r_{c}s_{c}/S_{\text{max}}$, where $r_c$ is the value of the Pearson correlation between disk triangulations, $s_{c}$ is the size of the community, i.e. the number of disks in that network, and $S_{\text{max}}$ is the size of the largest network \cite{Bassett15}. This factor quantifies the topological integrity of force networks by measuring the correlation between the topological distance along the network and the geometrical distance (Euclidean separation). Values of $g_{c}\rightarrow0$ indicate compact chains or linearly extended networks
with minimal to no branching. Conversely, $g_{c}\rightarrow1$ points towards highly branched networks with pronounced discontinuities (gaps) in the force transmission chains.

The mean network order parameter $L=N_{p}/N_{c}$, where $N_{p}$ and $N_c$ are the number of active particles forming chains and the number of chains, respectively. The parameter $L$ is interpreted as the number of particles in stress contact forming the average force chain. It roughly characterizes the spatial extent (in particle units, hence the order of the chain) of the average chain or network, indicating whether stress is localized in short chains or dispersed over extended regions in larger chains. In chains formed by mostly monodisperse disks, like in our experiments, $L$ can also be interpreted as the non-dimensional average chain length, $\bar{L}\approx L\,d_{s}$.

Together, these parameters provide complementary information: $g_c$ describes how stress is transmitted (linearly or compacted versus through branched networks), while $L$ describes the chain size or the number of particles participating in the total network. This dual perspective is essential for understanding segregation dynamics, as it clearly distinguishes between two fundamental aspects of force network evolution: topological reorganization (reflected in $g_c$) and spatial extension (reflected in $L$). By analyzing both metrics as functions of the intruder size ratio $R$, we can unravel the micromechanical origins of large-particle segregation and establish quantitative relationships between the force network architecture and intruder dynamics.

\section{Validation of previous segregation scaling law}
\label{sec:validation}

To validate our experimental results, we used Trewhela et al.'s scaling law for segregation \cite{Trewhela21a,Trewhela21b}. This validation yields the experimental constant $\mathcal{B}$ suited for our experimental conditions, which may differ slightly from those reported in the literature. Previous experiments and simulations reported values between 0 and 2, but such values may vary with friction, materiality, and the interstitial fluid between grains \cite{Trewhela21a,Trewhela21b,Kumawat25,Zhao25}.

Analysis of each experiment image sequence yields precise trajectories for all large particle intruders. Thus, the temporal evolution of the intruder vertical position $z^l$, as well as its velocity $w_{l}$ and segregation, can be fitted using the experimental scaling law proposed by Trewhela et al.\cite{Trewhela21a,Trewhela21b}, which describes the segregation rate in terms of the ordinary differential equation
\begin{equation}
    w_{l}=\frac{\text{d} z^{l}}{\text{d}t}=\mathcal{B}\frac{\partial p}{\partial z}\frac{\bar{\dot{\gamma}}d_{s}^{2}(R-1)}{p}.
    \label{eq:zl}
\end{equation}
In equation \ref{eq:zl}, $p=\rho \Phi g(h-z)$ is the lithostatic pressure around the intruder, while $g=9.81$ m/s$^2$ is the gravitational acceleration, $\Phi=0.7$ is the solids volume fraction, and $\rho=1040$ kg/m$^3$ is the intrinsic particle density. Notably, equation \ref{eq:zl} is separable and can be integrated, yielding
\begin{equation}
K^lt = \mathcal{C}d_s(z^l-z^l_0)+\frac{\Phi}{2}\left[(h-z^l_0)^2-(h-z^l)^2\right] = \mathcal Z^l(z^l),
\label{eq:zlint}
\end{equation}
Here $K^{l}=\mathcal{B}\bar{\dot{\gamma}}d_{s}^{2} (R-1)$ is the integration constant. The constant $\mathcal{C}=0.271$ is a fitted value for the experimental scaling law and is very similar to values previously reported \cite{Trewhela21a,Trewhela21b}. The $ode$ integration requires the initial position at $t=0$, $z_{0}^{l}$, which is taken from the intruder's initial position in each experiment. Equation \ref{eq:zlint} introduces the quadratic function for the trajectories, $\mathcal{Z}^{l}$, to be used in later trajectory collapse.

\begin{figure}[!ht]
\begin{center}
\SetLabels
\L (0.0*0.98) ($a$)\\
\L (0.5*0.98) ($b$)\\
\L (0.0*0.5) ($c$)\\
\L (0.58*0.69) \rotatebox{54}{\color{RoyalBlue}{$\mathcal{B}\bar{\dot{\gamma}} d_{s}^2=1.4\cdot10^{-5}$}}\\
\L (0.72*0.73) \rotatebox{28}{\color{RawSienna}{$\mathcal{B}\bar{\dot{\gamma}} d_{s}^2=5.4\cdot10^{-6}$}}\\
\L (0.78*0.6) $\mathcal{B}=0.8035$\\
\L (0.0*0.72) \rotatebox{90}{$z^{l}\quad$cm}\\
\L (0.25*0.49) $t\quad$s\\
\L (0.0*0.25) \rotatebox{90}{$z^{l}\quad$cm}\\
\L (0.47*0.0) $t\quad$s\\
\L (0.5*0.68) \rotatebox{90}{$\mathcal{K}^{l}\quad$m$^{2}$/s}\\
\L (0.69*0.495) $R-1$\\
\L (0.29*0.615) $d_{s}=5$ mm\\
\L (0.64*0.185) $d_{s}=8$ mm\\
\L (0.94*0.49) \rotatebox{90}{$\leftarrow R\rightarrow$}\\
\endSetLabels
\strut\AffixLabels{\includegraphics[width=5.5in]{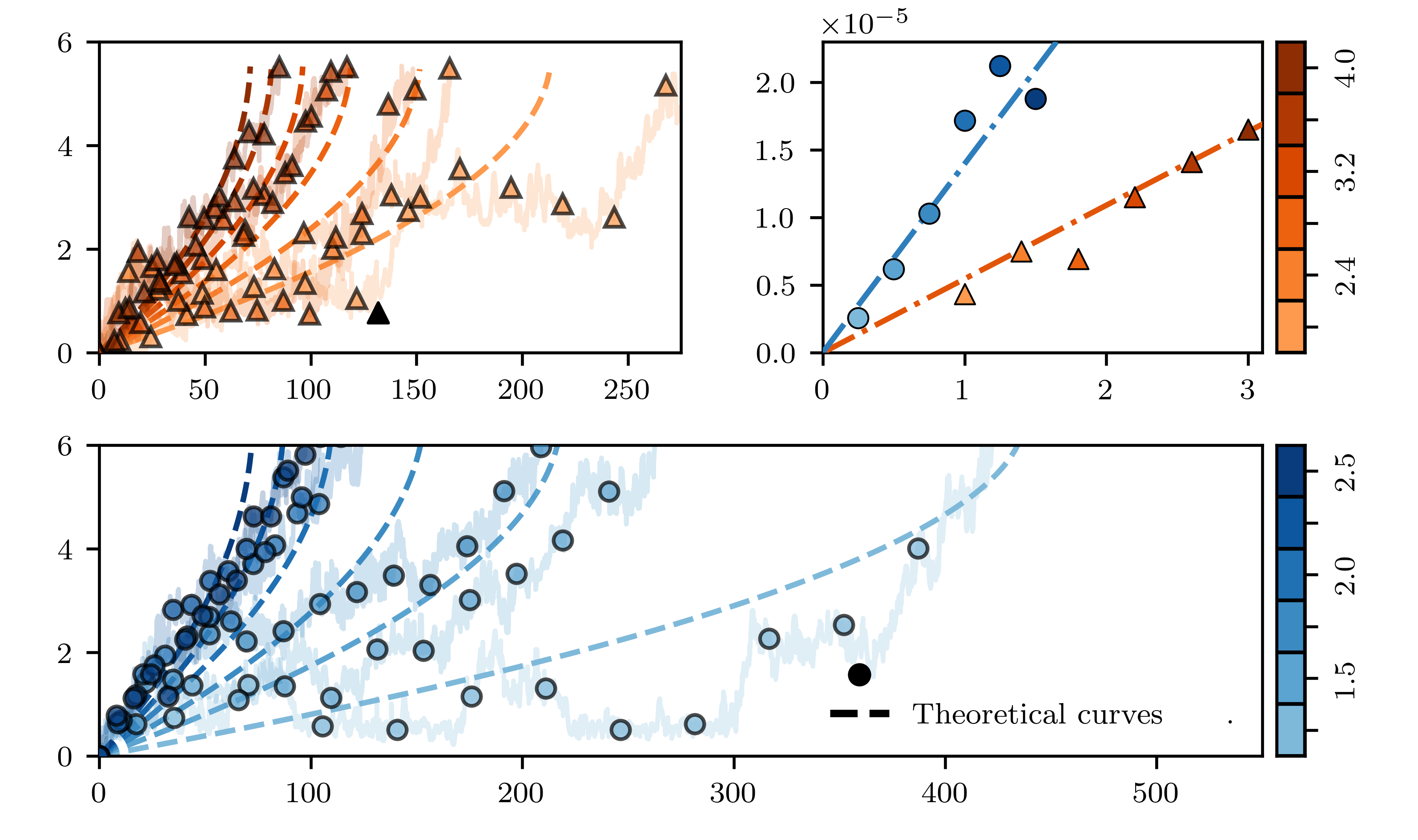}}
\end{center}
\caption{Large-particle position $z^{l}$ as a function of time $t$ for our experimental sets: ($a$) $d_s=5$ mm, ($c$) $d_s=8$ mm with intruders of $d_l=10, 12, 14, 16, 18 $ and $20$ mm. The dashed lines plot the theoretical curves in equation~\ref{eq:zlth} and the markers indicate the moving mean values each 25 positions. ($b$) Constants $K^{l}$ as a function of $(R-1)$. For each set of experimental results, a linear regression was performed to obtain $\mathcal{B}\bar{\dot{\gamma}}d_{s}^{2}$ and a constant $\mathcal{B}=0.8035$ for all our experiments.}
\label{fig:trajectories}
\end{figure}
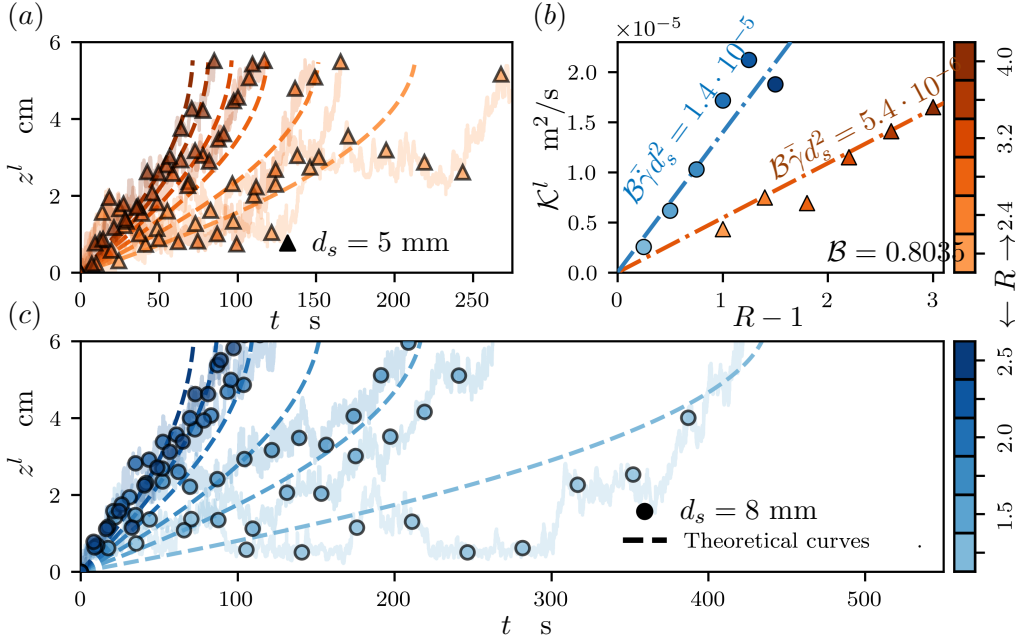

By solving $z_{l}$ in terms of $t$ from equation \ref{eq:zlint}, the theoretical trajectories can be obtained
\begin{equation}
    z^l = \frac{1}{\Phi}\left[\mathcal{C}d_s + \Phi h - \sqrt{\mathcal{C}^2d_s^2+2\mathcal{C}d_s\Phi(h-z^l_0)+\Phi^2(h-z^l_0)^2-2\Phi K^lt}\right].
    \label{eq:zlth}
\end{equation}
The obtained solutions are shown with our experimental trajectories in figure \ref{fig:trajectories}(\emph{a,c}). Our results confirm earlier findings: intruder segregation accelerates as the size ratio $R$ increases, with no plateau or maximum, consistent with prior work \cite{Trewhela21a,Trewhela21b} and distinct from other conditions [e.g. \citenum{Thomas18,Thornton12}]. The theoretical trajectories closely match our data (figure \ref{fig:trajectories}$a,c$), with $r^2$ values ranging from 0.79 to 0.97. Figure \ref{fig:trajectories}(\emph{c}) shows the integration constants $\mathcal{K}^{l}$ plotted against $R-1$. For both sets – indicated by $\bullet$ ($d_{s}=5$ mm) and $\blacktriangle$ ($d_{s}=8$ mm)– a linear regression with slope $\mathcal{B}=0.8035$ confirms data collapse. This $\mathcal{B}$ value matches previous reports and may guide future work with polyurethane disks.  

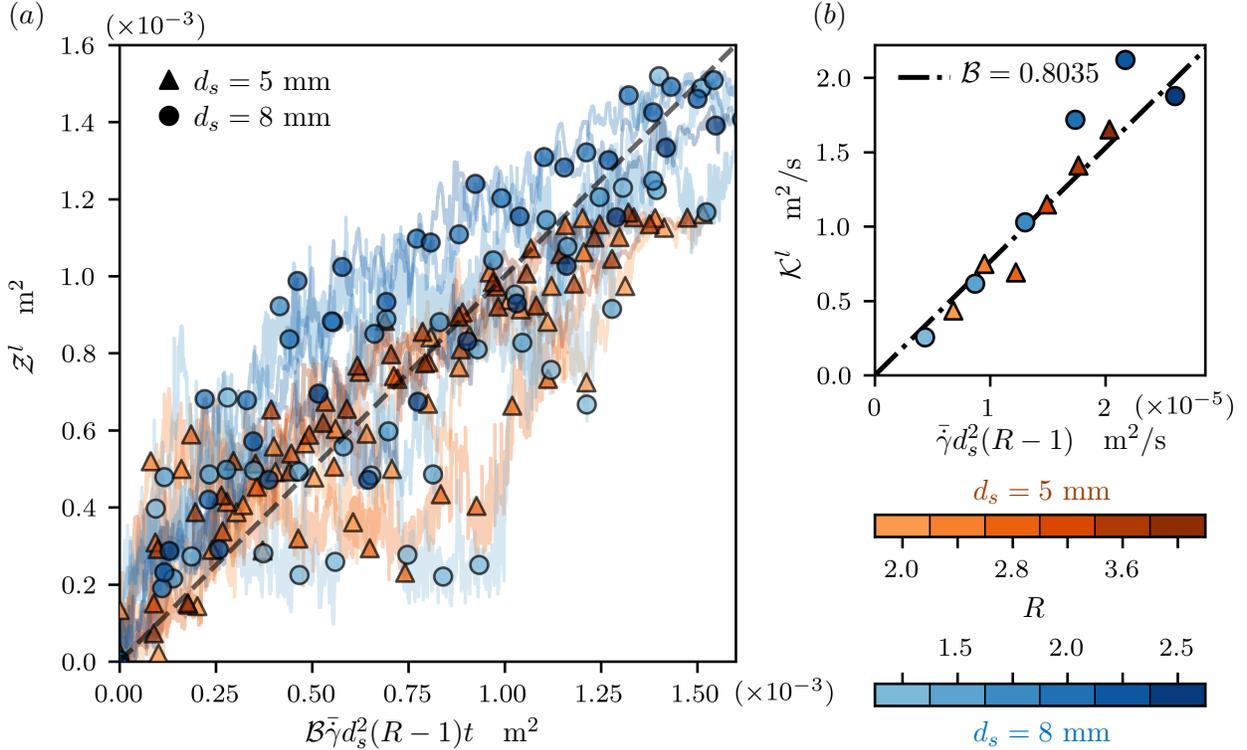
\begin{figure}[!ht]
\begin{center}
\SetLabels
\L (0.0*0.98) ($a$)\\
\L (0.65*0.98) ($b$)\\
\L (0.77*0.9) $\mathcal{B}=0.8035$\\
\L (0.0*0.5) \rotatebox{90}{$\mathcal{Z}^{l}\quad$m$^{2}$}\\
\L (0.24*0.0) $\mathcal{B}\bar{\dot{\gamma}}d_{s}^{2}(R-1)t\quad$m$^{2}$\\
\L (0.62*0.63) \rotatebox{90}{$\mathcal{K}^{l}\quad$m$^{2}$/s}\\
\L (0.75*0.4) $\bar{\dot{\gamma}}d_{s}^{2}(R-1)\quad$m$^{2}$/s\\
\L (0.15*0.89) $d_{s}=5$ mm\\
\L (0.15*0.84) $d_{s}=8$ mm\\
\L (0.82*0.17) $R$\\
\L (0.78*0.33) \textcolor{RawSienna}{$d_{s}=5$ mm}\\
\L (0.78*0.00) \textcolor{RoyalBlue}{$d_{s}=8$ mm}\\
\endSetLabels
\strut\AffixLabels{\includegraphics[width=\textwidth]{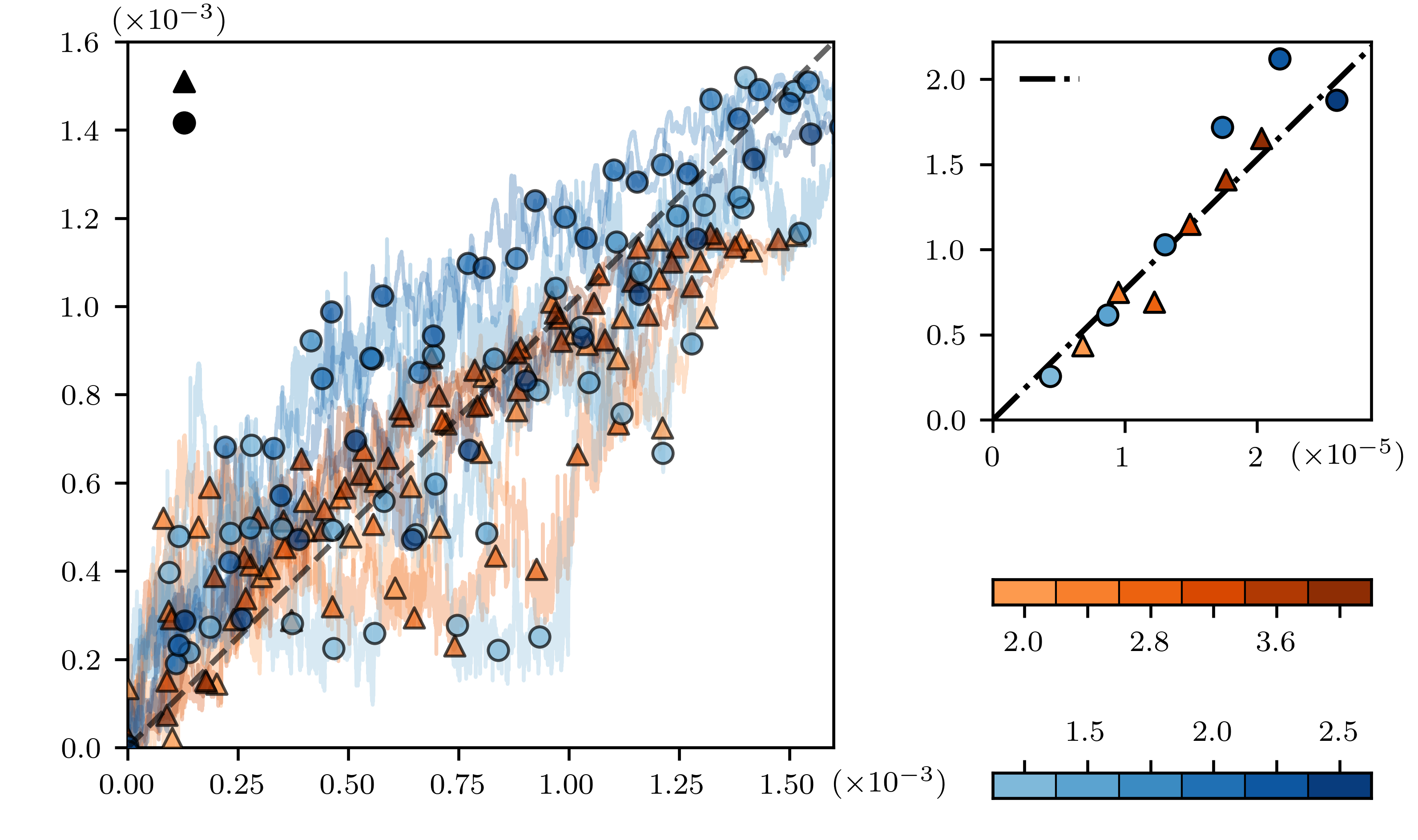}}
\end{center}
\caption{
($a$) All intruder trajectories collapsed onto the identity dashed line, given by the parametrized depth $Z^l$ and time $\mathcal{B} \dot{\gamma} d_s^2 (R-1)t$, according to equation~\ref{eq:zlint}. ($b$) All fitted constants $K^l$ for our experiments as a function of $\bar{\dot{\gamma}} d_s^2 (R-1)$. In the plot, the dashed line corresponds to the segregation constant $\mathcal{B} = 0.8035$ determined by a least-squares fit of our data.}
\label{fig:zlcollapse}
\end{figure}

Figure~\ref{fig:zlcollapse}(\emph{a}) demonstrates that $\mathcal{B}$ enables all experimental data to collapse onto a single plot. This is accomplished using the parametrized trajectories $\mathcal{Z}^{l}$ (equation \ref{eq:zlint}) and the time $\mathcal{B} \dot{\gamma} d_s^2 (R-1)t$, plotting the integration constants $\mathcal{K}^{l}$ against $\bar{\dot{\gamma}}d_{s}^{2}(R-1)$. The result is an excellent data collapse matching the segregation constant $\mathcal{B}=0.8035$. The data collapse validates our experimental scaling law, which provides a robust parametrization despite some variability, and the observed trend aligns well with theoretical predictions.

\section{Results}
\label{sec:results}

The role of the stress networks on intruder segregation was assessed by calculating the statistics of both the gap factor $g_{c}$ and the mean network order parameter $L$. For each parameter, we fitted a probability distribution function ($PDF$) and determined a relationship between the corresponding $PDF$ defining parameter and our experimental control parameters.

\subsection{Gap-factor distribution}

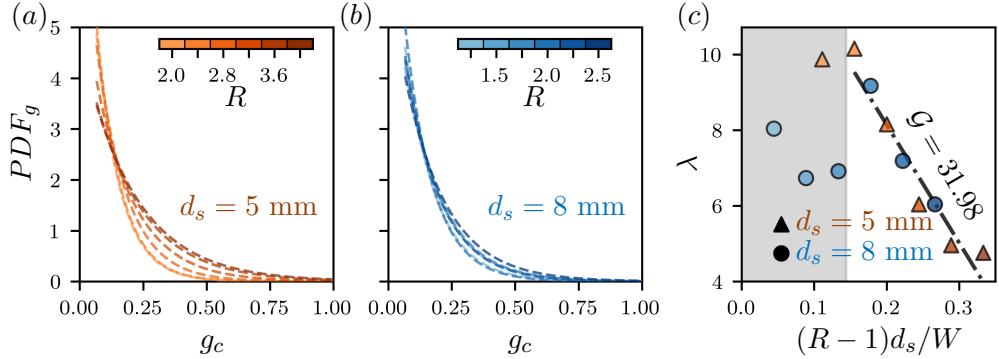
\begin{figure}[!ht]
\begin{center}
\SetLabels
\L (0.0*0.98) ($a$)\\
\L (0.33*0.98) ($b$)\\
\L (0.67*0.98) ($c$)\\
\L (0.9*0.7) \rotatebox{-60}{$\mathcal{G}=31.98$}\\
\L (0.0*0.5) \rotatebox{90}{$PDF_g$}\\
\L (0.19*0.0) $g_{c}$\\
\L (0.50*0.0) $g_{c}$\\
\L (0.67*0.53) \rotatebox{90}{$\lambda$}\\
\L (0.79*0.0) $(R-1)d_{s}/W$\\
\L (0.215*0.73) $R$\\
\L (0.515*0.73) $R$\\
\L (0.79*0.355) \textcolor{RawSienna}{$d_{s}=5$ mm}\\
\L (0.79*0.265) \textcolor{RoyalBlue}{$d_{s}=8$ mm}\\
\L (0.17*0.4) \textcolor{RawSienna}{$d_{s}=5$ mm}\\
\L (0.48*0.4) \textcolor{RoyalBlue}{$d_{s}=8$ mm}\\
\endSetLabels
\strut\AffixLabels{\includegraphics[width=5.2in]{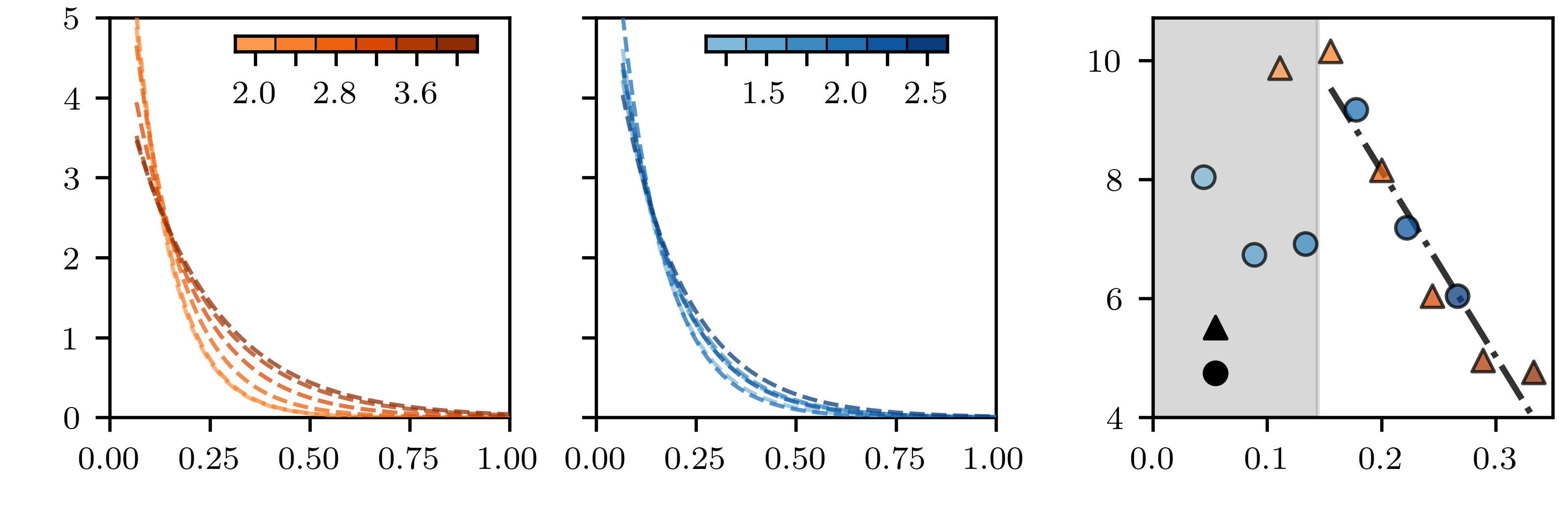}}
\end{center}
\caption{($a-b$) Probability distribution functions of the gap factor $g_c$ for varying size ratios $R$ and for both experimental sets with $d_s=5$ and 8 mm. ($c$) Parameter of the exponential distribution $\lambda$ (equation \ref{eq:pdf_gc}) as a function of $(R-1)d_{s}/W$, given by equation \ref{eq:gc_param}. The dash-dot line represents the slope $\mathcal{G}=31.98$ resulting from the linear regression in \ref{eq:gc_param}.}
\label{fig:gc}
\end{figure}

Since network branching decays exponentially as the network develops, the probability of having a network with a certain value of gap factor $g_{c}$ follows an exponential distribution, i.e.
\begin{equation}
    PDF_{g}=\lambda e^{-\lambda g_{c}},
    \label{eq:pdf_gc}
\end{equation}
with $\lambda$ the parameter of the distribution. Figure \ref{fig:gc}(\emph{a,b}) depicts the $PDF_{g}$ functions for the experimental data sets, $d_{s}=5$ mm and 8 mm, respectively. We observe from both plots that the probability of obtaining low $g_{c}$ (compact) networks is high, and that probability decreases with $R$ at the intersection of the curves where $g_c\lessapprox0.125$. Conversely, we obtain low probabilities of developing high-$g_c$-valued networks (chains), but these probabilities now increase with $R$. These results are key to determining the dependence of the distribution parameter $\lambda$. The plots in figure \ref{fig:gc} suggest a linear relationship with the size ratio $R$. Also, the gap factor $g_{c}$ should decline exponentially with the chain extent, determined by the particle diameter $d\approx d_{s}$, thus yielding the integral relationship for the probabilities
\begin{equation}
\lambda = -\mathcal{G}(R-1) \frac{d_s}{W} + n_{g},
\label{eq:gc_param}
\end{equation}
with $\mathcal{G}=31.98$ and $n_{g}=14.61$ the linearly regressed parameters. 

In figure \ref{fig:gc}(\emph{c}), we plot the experimental $\lambda$ values as a function of $(R-1) \frac{d_s}{W}$, as expressed in equation \ref{eq:gc_param}. It is worth noting that the fit is only valid for experiments with size ratios $R\gtrapprox2$. This validity is due to the fact that for smaller size ratios $R<2$ (see the shaded area in figure \ref{fig:gc}\emph{c}), the gap factor distribution is dominated by compact linear chains, with limited branching and a lesser role of the intruder in determining the stress network characteristics. As $R$ approaches and exceeds 2, $\lambda$ decreases systematically with $(R-1)d_s/W$. Hence, the probability distribution decays more slowly with increasing $g_c$, and the likelihood of large gap factors rises. Therefore, due to their greater extent ($d_{l}>d_{s}$) and self-induced size heterogeneity, larger intruders promote stress networks with more branches, characterized by pronounced internal gaps and discontinuities.

\subsection{Mean network order}

To model the statistics of the mean network order parameter $L$, or the average number of particles forming a force network, we chose a power-law distribution
\begin{equation}
    PDF_{L}=\beta L^{-\alpha_{L}},
    \label{eq:pdfL}
\end{equation}
where $\alpha$ and $\beta$ are the power law defining parameters. These parameters can be related by integrating equation \ref{eq:pdfL} yielding $\beta=(\alpha-1)L_{\text{min}}^{(\alpha-1)}$ with $L_{\text{min}}=3$, the minimum mean network order, i.e. a single chain $N_c=1$ of three particles $N_p=3$. With the latter relationship, the power law distribution can be described by a single independent parameter, i.e. $\alpha$.

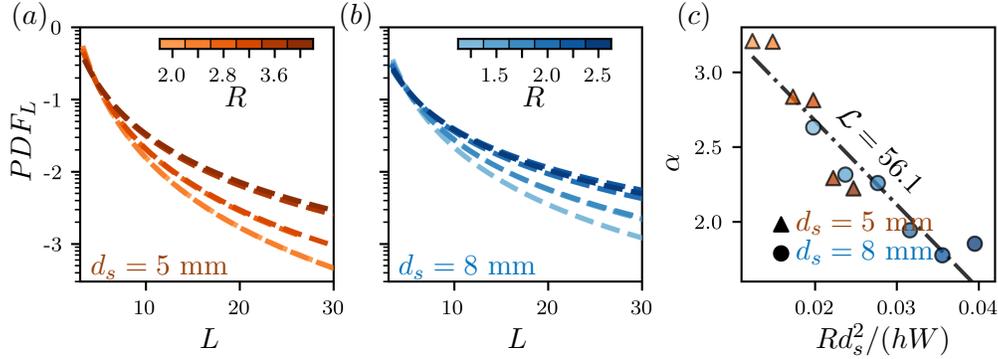
\begin{figure}[!ht]
\begin{center}
\SetLabels
\L (0.0*0.98) ($a$)\\
\L (0.33*0.98) ($b$)\\
\L (0.67*0.98) ($c$)\\
\L (0.83*0.7) \rotatebox{-46}{$\mathcal{L}=56.1$}\\
\L (0.0*0.5) \rotatebox{90}{$PDF_L$}\\
\L (0.19*0.0) $L$\\
\L (0.50*0.0) $L$\\
\L (0.66*0.53) \rotatebox{90}{$\alpha$}\\
\L (0.81*0.0) $Rd^{2}_{s}/(hW)$\\
\L (0.215*0.73) $R$\\
\L (0.515*0.73) $R$\\
\L (0.79*0.355) \textcolor{RawSienna}{$d_{s}=5$ mm}\\
\L (0.79*0.265) \textcolor{RoyalBlue}{$d_{s}=8$ mm}\\
\L (0.08*0.22) \textcolor{RawSienna}{$d_{s}=5$ mm}\\
\L (0.39*0.22) \textcolor{RoyalBlue}{$d_{s}=8$ mm}\\
\endSetLabels
\strut\AffixLabels{\includegraphics[width=5.2in]{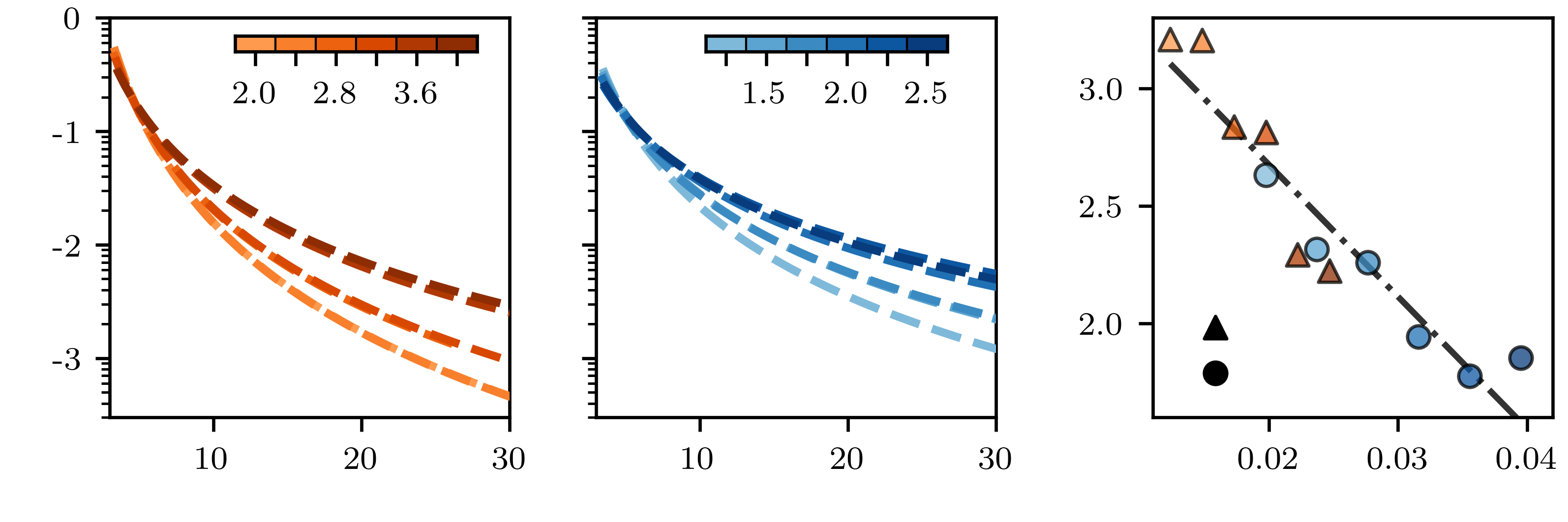}}
\end{center}
\caption{($a-b$) Probability distribution functions of the mean network order $L$ for varying size ratios $R$ and for both experimental sets with $d_s=5$ and 8 mm. ($c$) Defining parameter of the power-law distribution $\alpha$ (equation \ref{eq:pdfL}) as a function of $Rd^{2}_{s}/(hW)$, given by equation \ref{eq:Lparam}. The dash-dot line represents the slope $\mathcal{L}=56.1$ obtained via the linear regression of \ref{eq:Lparam}.}
\label{fig:L}
\end{figure}

The power law distribution in equation \ref{eq:pdfL} successfully fits the experimental data for the $d_{s}=5$ and 8 mm sets in figure \ref{fig:L} (\emph{a,b}). We observe a similar trend to that of figure \ref{fig:gc}(\emph{a,b}); probabilities of encountering larger $L$ values increase with $R$ for both sets. As mentioned earlier, the self-induced size heterogeneity of the intruder not only produces larger gap factors but also larger orders in stress networks, thereby leading to more widely distributed networks. This observation becomes evident by picturing the effect of a very large intruder in the cell. Larger size implies that the intruder will be in contact with more small particles, and, consequently, sidewall stresses are more likely to be transmitted than in the case of smaller intruders.

As done for $\lambda$, it was possible to parameterize the power-law parameter $\alpha$ as a function of our experimental control parameters. A function was drawn from the previously detailed size ratio $R$ dependence, with the notion that a full single network should be made of all the particles; then, a non-dimensional area $d_{s}^{2}/(hW)$ yielded the scaling

\begin{equation}
\alpha = -\mathcal{L}\,R\,\frac{d_s^2}{hW} + n_L,
\label{eq:Lparam}
\end{equation}
where $\mathcal{L}=56.1$ and $n_L=3.797$ are the linearly regressed parameters. In figure \ref{fig:L}(\emph{c}), we plot the experimental $\alpha$ values as a function of $R\,d_s^2/(hW)$, showing a good adjustment of the proposed parametrization. The proposed relation \ref{eq:Lparam} suggests that the chain main order $L$ is controlled by the large-small particle surface $Rd_{s}^2=d_{l}d_{s}$. Consequently, larger intruders generate larger stress networks, whereas smaller intruders produce shorter chains that are confined to their vicinity. The scaling is also consistent with the monodisperse case $R=1$, in which stress networks still form and a defined value for $\alpha$ can be obtained for each $d_{s}$.

\subsection{Conditional probability for segregation}

After characterizing and parameterizing the statistics of stress networks, it is important to relate them to intruder segregation. With that goal in mind, the use of conditional probabilities has enabled the correlation of segregation with other events, such as rotation or local changes in particle concentration \cite{vanderVaart15,Trewhela21b}. Qualitatively, in our experiments, we observed that large particle segregation was associated with the formation of large stress networks and their durable transmission to large particles. This observation is quite intuitive, but it is not new, as it can be directly tied to the original description of the squeeze expulsion mechanism \cite{Savage88}, in experiments \cite{Trewhela21b} and  simulations \cite{Jing17,Staron18}. Compared to previous experimental studies, we have access to stress network lengths and can effectively assess whether a large particle segregates due to the formation of large stress chains.

\begin{figure}[!ht]
\begin{center}
\SetLabels
\L (0.0*0.97) ($a$)\\
\L (0.0*0.73) ($b$)\\
\L (0.0*0.49) ($c$)\\
\L (0.0*0.25) ($d$)\\
\L (0.67*0.97) ($e$)\\
\L (0.67*0.73) ($f$)\\
\L (0.67*0.49) ($g$)\\
\L (0.67*0.25) ($h$)\\
\L (0.0*0.87) \rotatebox{90}{$L$}\\
\L (0.0*0.63) \rotatebox{90}{$L$}\\
\L (0.0*0.39) \rotatebox{90}{$L$}\\
\L (0.0*0.15) \rotatebox{90}{$L$}\\
\L (0.30*0.0) $t$\quad s\\
\L (0.65*0.83) \rotatebox{90}{$|w_{l}|$\quad m/s}\\
\L (0.65*0.59) \rotatebox{90}{$|w_{l}|$\quad m/s}\\
\L (0.65*0.35) \rotatebox{90}{$|w_{l}|$\quad m/s}\\
\L (0.65*0.11) \rotatebox{90}{$|w_{l}|$\quad m/s}\\
\L (0.85*0.0) $L$\\
\L (0.82*0.94) \textcolor{white}{$\zeta_c=-0.684$}\\
\L (0.82*0.70) \textcolor{white}{$\zeta_c=-0.280$}\\
\L (0.82*0.31) \textcolor{white}{$\zeta_c=0.228$}\\
\L (0.82*0.07) \textcolor{white}{$\zeta_c=0.275$}\\
\L (0.08*0.22) \textcolor{RawSienna}{$d_{s}=5$ mm, $R=4$}\\
\L (0.08*0.46) \textcolor{RoyalBlue}{$d_{s}=8$ mm, $R=2.5$}\\
\L (0.08*0.70) \textcolor{RawSienna}{$d_{s}=5$ mm, $R=2$}\\
\L (0.08*0.94) \textcolor{RoyalBlue}{$d_{s}=8$ mm, $R=1.5$}\\
\endSetLabels
\strut\AffixLabels{\includegraphics[width=5.2in]{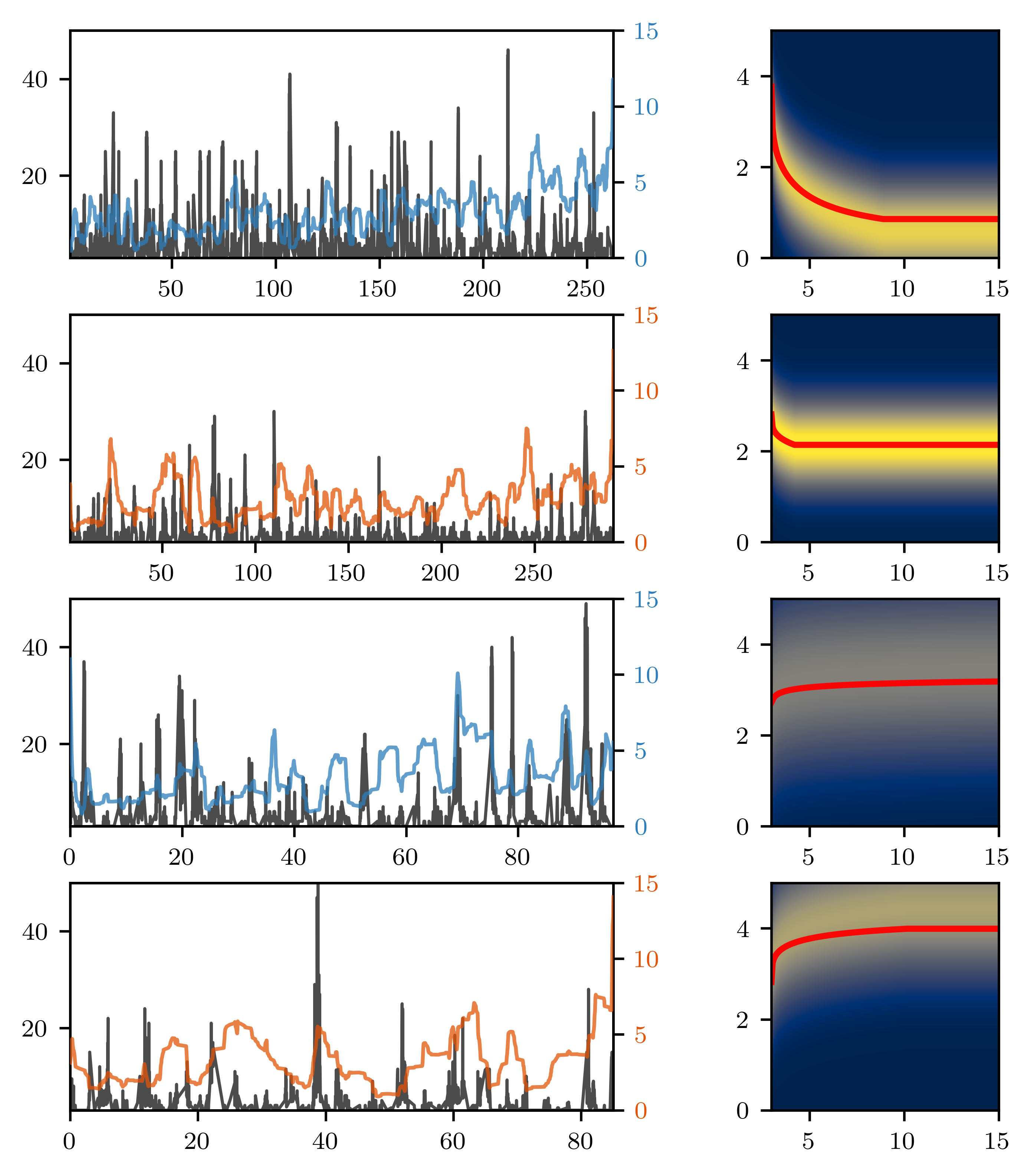}}
\end{center}
\caption{($a-d$) Time series of the intruder vertical velocity absolute value $|w_{l}|$ (in color) and the mean network order $L$ (in gray) for various size ratios $R$. ($e-h$) Conditional probability for segregation $|w_{l}|$ given a mean network order $L$, $P(w_{l}|L)$ for the corresponding size ratio $R$ cases of panels $a-d$. Probabilities range from 0.0 (dark blue) to 0.6 (yellow). For each colormap of $P(w_{l}|L)$, the corresponding copula correlation coefficients $\zeta_c=\lbrace-0.684, -0.280, 0.228,0.275\rbrace$ were calculated.}
\label{fig:w_L}
\end{figure}
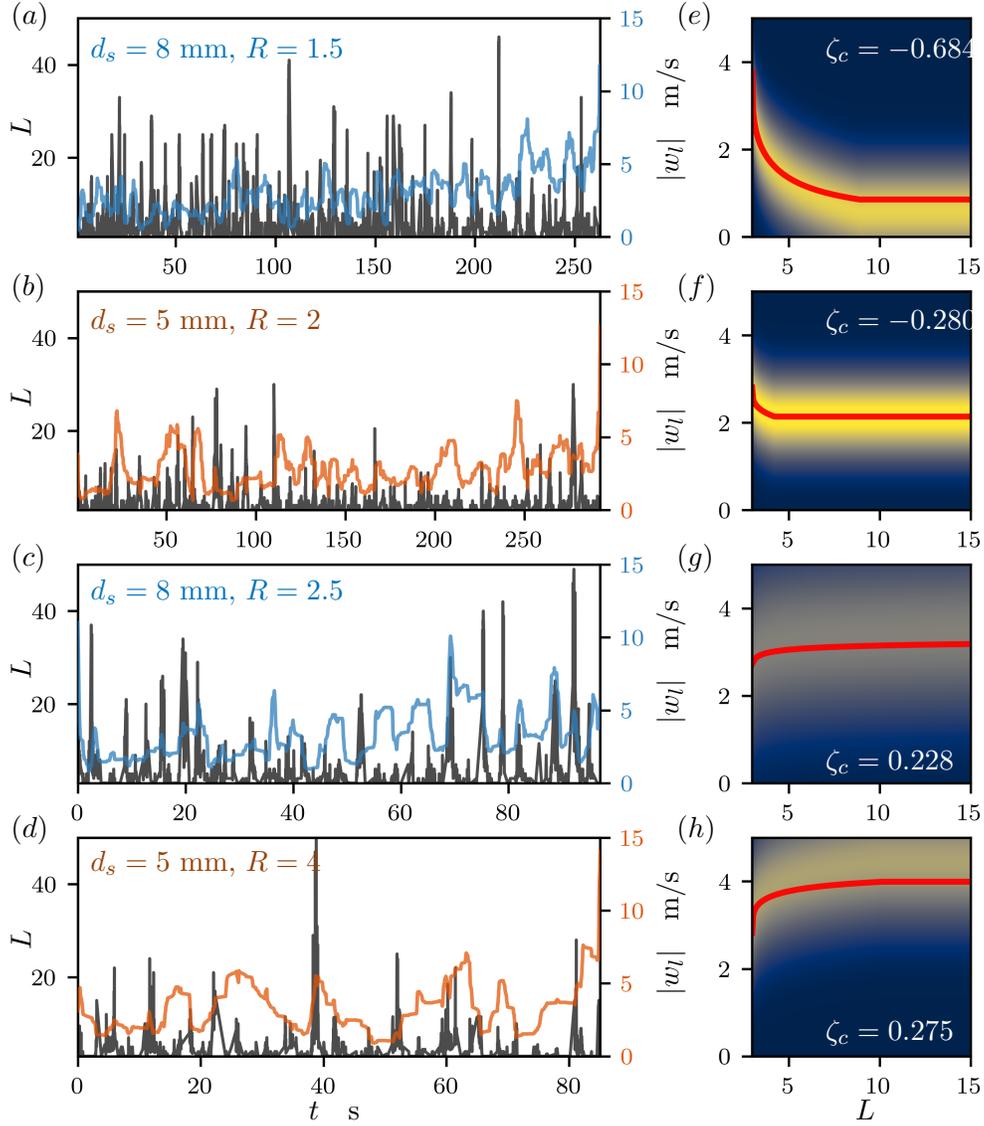

To model the full bivariate dependence between the intruder segregation rate $|w_{l}|$ and the mean chain order $L$ without imposing statistical restrictions on the variables, we employed a Gaussian copula \cite{Genest07}. Unlike standard multivariate Gaussian models, which require variables to be normally distributed, the Gaussian copula only assumes a Gaussian dependent structure while allowing the marginal distributions of $|w_{l}|$ and $L$ to take arbitrary functional forms, as introduced in equation \ref{eq:pdfL} for $L$. With this flexibility, a degree of coupling or copula coefficient $\zeta_c$ rises as the key parameter to evaluate how strongly the variables are coupled. Values of $\zeta_c$ range from -1 to 1, indicating, in our case, that larger chains correlate with slower and faster segregation rates, respectively.

Figure \ref{fig:w_L}(\emph{a–d}) displays the moving average of the intruder segregation velocity $|w_{l}|$ and the mean order of the stress network $L$ across four representative experiments with different size ratios $R=\lbrace 1.5, 2.0, 2.5, 4.0 \rbrace$. For larger size ratios ($R>2$), we observe that an increase in intruder velocity coincides with the formation of more extensive stress networks, indicating that strong force chains enhance the mobility of large intruders. However, for $R=1.5$, there is no clear coupling between network size and velocity; in these cases, even well-developed stress networks do not translate into efficient segregation. Thus, the size ratio $R$ appears to modulate how stress is transmitted and how force networks facilitate the self-driven mechanisms that enable large-particle segregation.

The conditional probabilities help us better understand the observed correlation. In figure \ref{fig:w_L}(e-h), velocity-chain length coupling transitions clearly as $R$ increases. For smaller size ratios (figure \ref{fig:w_L}e-f), intruder velocity decreases as chain length increases, leading to negative copula correlation ($\zeta_c<0$). This matches force chain-mediated drag: longer chains transfer more resistive force from the medium, slowing the intruder's movement \cite{Bassett15,Fang20}. For small intruders, short-chain networks—which cause faster segregation—are rare. More common, larger networks dominate; these create a compact bulk that impedes segregation. This paradox is resolved when recognizing that small intruder segregation is not aided by long stress networks. Instead, shorter stress networks provide less resistance and enough support for segregation. 

As the size ratio approaches $R\approx2$, the system enters a transitional regime in which the correlation weakens. This crossover aligns with experimental scaling laws for segregation velocity, which identify $R\approx2$ as a characteristic threshold separating gravity-dominated and kinematics-dominated segregation mechanisms \cite{Guillard16,Trewhela21b}.

Finally, for larger size ratios $R>2$ (figure \ref{fig:w_L}\emph{g-h}), the conditional mean value becomes nearly flat, and the copula correlation approaches zero or turns slightly positive. In this regime, the intruder velocity is largely independent of the stress network length, indicating that the particle has transitioned from a caged state, where mobility is governed by the surrounding force network, to an expansion or dilation dominated state \cite{Rosato87, Trewhela21b}. Here, the aforementioned paradox inverts; extended networks result in faster segregation. These networks exhibit pronounced branching (high $g_c$), which translates into bulk reorganization rather than large-scale resistance, signifying the previously-described dilation mechanism.

\section{Conclusion}

We conducted experiments in a bidimensional shear cell filled with polyurethane disks to study particle-size segregation of intruders larger than the surrounding grains. We used photoelasticity to visualize stress transmission and network formation as the large intruder segregates vertically through the bulk. We extracted the intruders' positions and successfully validated the scaling law of Trewhela et al. \cite{Trewhela21a} by collapsing their trajectories onto the theoretical curve defined by a segregation coefficient $\mathcal{B}=0.8035$, finding a useful parameter for future research on segregation of photoelastic particles. The experimental stress networks were characterized by their gap factor $g_{c}$ and mean network order $L$ time series. Both $g_c$ and $L$ were described by their \emph{PDF}s. Their distribution parameters, $\lambda$ and $\alpha$, depend on the size ratio $R$, indicating that stress architecture and distribution are intruder-size dependent. For large $R$ values, the probability of developing high gap factor networks increases with respect to smaller $R$ experiments. The observation is similar for $L$; the likelihood of extended stress networks with large $L$ values increases with $R$.

Finally, for larger size ratios $R>2$ (figure \ref{fig:w_L}\emph{g-h}), the conditional mean value becomes nearly flat, and the copula correlation approaches zero or turns slightly positive. In this regime, the intruder velocity is largely independent of the stress network length, indicating that the particle has transitioned from a caged state, where mobility is governed by the surrounding force network, to an expansion or dilation-dominated state \cite{Rosato87, Trewhela21b}. The aforementioned paradox inverts in this case: extended networks lead to faster segregation. These networks exhibit pronounced branching (i.e., high $g_c$), leading to bulk reorganization rather than large-scale resistance, consistent with the previously described dilation mechanism. 

\begin{acknowledgments}
The authors are grateful to Pierre-Yves Lagrée and Anaïs Abramian for hosting A. N. in a short internship at Institut Jean Le Rond $\partial$'Alembert and their constructive comments on the experimental setup and methods. T. T. is grateful for the technical support of Bob de Graffenried and Michel Teuscher in the experimental setup development.

This research received support from Dirección de Investigación y Doctorados (DIID) of Universidad Adolfo Ibañez through Formación de Estudiantes en Investigación. A. N. is grateful for the support given by Instituto Francés Chile and Dirección de Investigación y Postgrados Académicos (DIPA) of Facultad de Ingeniería y Ciencias through their short internship scholarship program. T. T. and L. G. received support from Agencia Nacional de Investigación y Desarrollo (ANID) via FONDECYT Iniciación 11240630 and FONDECYT Regular 1260956, respectively.
\end{acknowledgments}

\bibliography{sample}

\end{document}